\documentclass[12pt]{article}
\usepackage{latexsym,amsmath,amssymb,amsbsy,graphicx}
\usepackage[cp1251]{inputenc}
\usepackage[T2A]{fontenc}
\usepackage[russian]{babel}
\usepackage[space]{cite} % If exist.

\makeatletter\def\@biblabel#1{\hfill#1.}\makeatother

\allowdisplaybreaks
\begin {document}
% ======================================

\noindent\begin{minipage}{\textwidth}
	\begin{center}
		
		% {\large ФИЗИКА ... \small(рубрика прописными)}\\[1cc]
		% Рубрика обычно указывается редактором.
		
		{\Large{Создание фотоннокристаллических структур с произвольными спектральными особенностями}}\\[9pt]
		
		{\large С.\,Е.\,Свяховский$^{a}$, Н.\,И.\, Пышков}\\[6pt]
		
		% Цифровые индексы указываются, если у авторов разные почтовые адреса;
		% буквенные -- если авторов несколько, для всех авторов, отвечающих на корреспонденцию.
		% Ниже и в сведениях на английском -- аналогично.
		
		\parbox{.96\textwidth}{\centering\small\it
			Московский государственный университет имени М.\,В.~Ломоносова, физический факультет,
			кафедра общей физики.
			Россия, 119991, Москва, Ленинские горы, д.~1, стр.~2.\\
			
			\ E-mail: $^a$sse@shg.ru,
		}\\[1cc] % не более 2 e-mail адресов в шапке статьи!!!!
		
		\parbox{.96\textwidth}{\centering\small Статья поступила --.--.2023, подписана в печать --.--.2023.} % Даты указываются редактором.
	\end{center}
	
	{\parindent5mm Мы представляем метод создания фотоннокристаллических структур, оптический спектр коэффициента отражения которых удовлетворяет заданной форме и имеет заранее известные особенности. Разработан алгоритм построения фотоннокристаллической структуры, выполнено численное моделирование её спектров отражения и создан экспериментальный образец фотонного кристалла, имеющего спектральный отклик, соответствующий заданной форме.\vspace{2pt}\par}

	\textit{Ключевые слова}: фотонные кристаллы, спектроскопия, обратное проектирование, пористый кремний.\vspace{1pt}\par
	
	\small УДК: 535.44 \vspace{1pt}\par
	\vspace{1pt}\par
\end{minipage}

\section*{Введение}
\mbox{}\vspace{-\baselineskip}

Фотонные кристаллы (ФК) --- структуры с пространственной периодичностью показателя преломления на масштабах, сопоставимых с длиной волны видимого излучения \cite{joannopoulos1997photonic2}. Главное свойство ФК --- существование области частот, в которой распространение света внутри кристалла невозможно, так называемой фотонной запрещённой зоны (ФЗЗ) \cite{yablonovitch1994photonic,joannopoulos1997photonic}. Фотонный кристалл может иметь медленно меняющийся период, а также его структура может состоять из суммы множества периодических функций \cite{singh2020multi}, при этом в его спектре могут обнаруживаться несколько ФЗЗ, каждая из которых соответствует своей периодической функции \cite{vyunishev2017quasiperiodic}, при этом благодаря наличию ФЗЗ эти структуры также именуются фотонными кристаллами.

 Наличие ФЗЗ позволяет эффективно управлять светом, что определяет широкую сферу применения фотонных кристаллов во многих областях оптики: они используются для усиления комбинационного рассеяния\cite{benabid2002stimulated, ashurov2020photonic}, нелинейно-оптических эффектов\cite{mcgurn2015nonlinear, arie2010periodic}, таких как генерация второй\cite{martorell1997second,fedyanin2005nonlinear} и третьей\cite{markowicz2004dramatic,martemyanov2004third} оптических гармоник, сверхбыстрого оптического переключения \cite{mazurenko2003ultrafast,hache2000ultrafast} и множества других применений, благодаря которым последние 30 лет фотонные кристаллы являются объектами интенсивных исследований.

Исходя из обширной сферы применения фотоннокристаллических материалов, возникает задача проектирования фотонных кристаллов с заданным положением ФЗЗ, имеющие заданные спектральные особенности. Например, для усиления оптических эффектов или спектральной фильтрации требуется добавить в спектр микрорезонаторную моду, или при проектировании делителей пучка требуется добавить плоский участок с заданным коэффициентом отражения.

 Периодическая модуляция показателя преломления в пространстве в одном направлении позволяет получить фотонную запрещённую зону. В спектральном отклике ФК, а именно в зависимости коэффициента отражения $R$ от длины волны $\lambda$, фотонная запрещённая зона будет иметь вид участка с высоким коэффициентом отражения, спектральная ширина и высота этого максимума будет определяться амплитудой модуляции показателя преломления и суммарной толщиной ФК. В том случае, когда пространственная модуляция показателя преломления происходит по гармоническому закону, в спектре ФК имеется лишь одна запрещённая зона без зон высших порядков, а ширина ФЗЗ определяется амплитудой гармонической функции модуляции. При достаточно малой амплитуде пространственной модуляции показателя преломления спектр отражения имеет вид узкого пика. Представляется возможным задать произвольную функцию спектрального отклика $f(\lambda)$, которая может быть аппроксимирована суммой этих спектральных пиков. В данной работе мы демонстрируем алгоритм разложения произвольной функции $f(\lambda)$ по пространственным гармоникам фотонного кристалла и проверяем его экспериментально.

\section{Выбор желаемой функции спектрального отклика}

\begin{figure}[tbh!]
	\centerline{\includegraphics[width=0.7\linewidth]{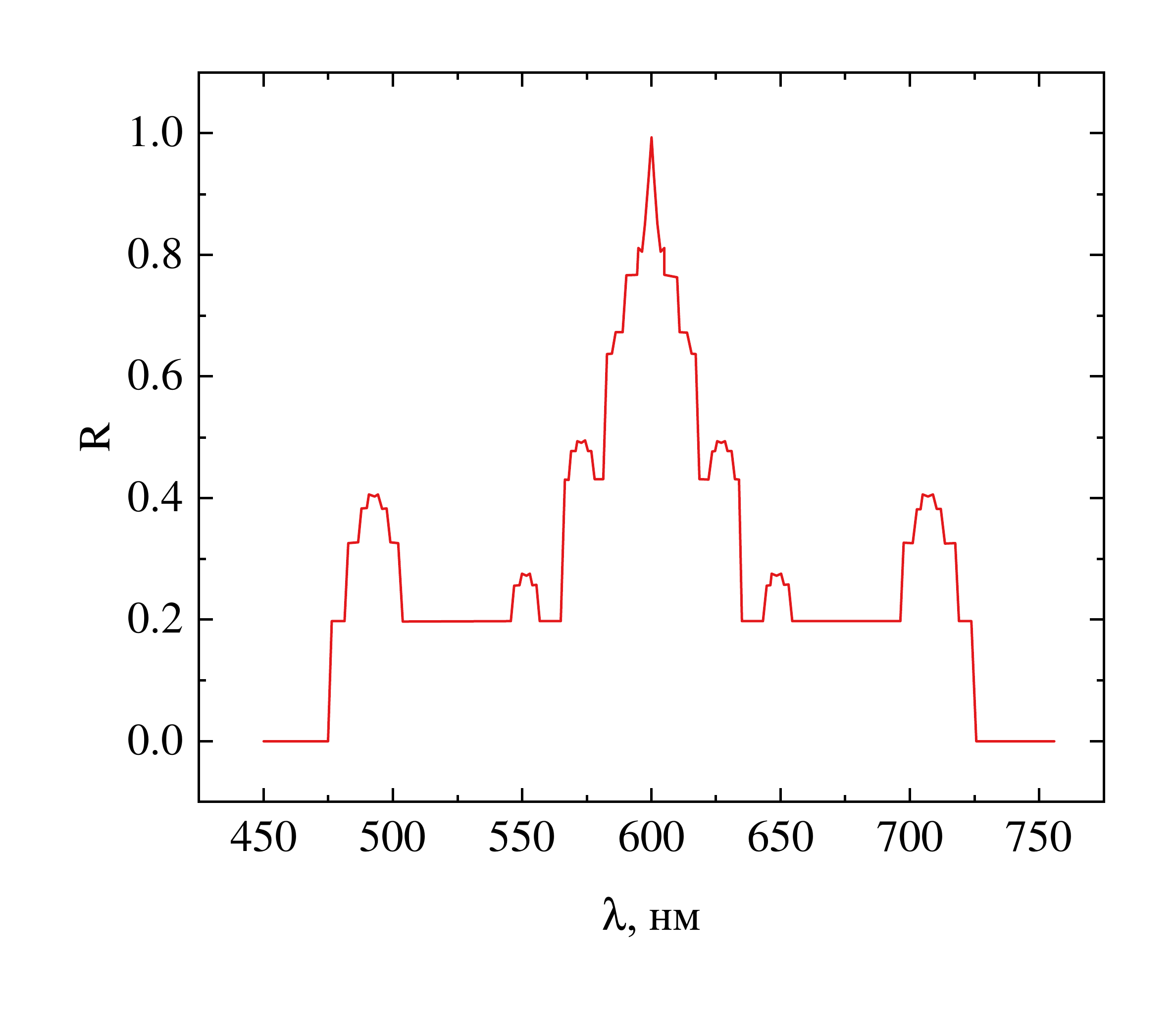}}
	\caption{Желаемая функция спектрального отклика} \label{OriginalFunction}
\end{figure}

В качестве произвольной желаемой функции спектрального отклика была выбрана функция, график которой представлен на рис. \ref{OriginalFunction}. Функция задавалась по координатам точек, её задание в виде формулы выходит за рамки данной работы.

Выбор именно этой желаемой функции был сделан из соображений наличия на ней спектральных особенностей, которые являются показательными и достаточно трудными для воспроизведения, а именно:
\begin{enumerate}
	\item П-образные участки, амплитуда которых отличается от 1. Известно, что для фотонного кристалла очень легко подобрать параметры, при котором его спектр отражения будет иметь П-образную форму с коэффициентом отражения, близким к 1, а именно такую форму будет иметь обычная ФЗЗ периодического ФК. Случай, когда коэффициент отражения П-образного участка имеет отличную от 1 амплитуду, является нетривиальным.
	\item Наклонные участки. Наличие наклонных участков демонстрирует возможность создавать фотоннокристаллические структуры, плавно меняющие спектр входного сигнала, что актуально при исследованиях фемтосекундных оптических импульсов.
	\item Спектральные особенности разной степени детализации. На используемой нами функции имеются участки как крупного масштаба, так и более детализованные. Воспроизведение мелких деталей далее было использовано в работе как оценка качества численного моделирования.
	\item Повторяющиеся особенности разного масштаба. Функция имеет особенности, похожие по форме, но различные по масштабу, например, максимумы на длинах волн 490 и 560 нм. Воспроизведение таких спектральных особенностей позволит сделать вывод о возможности применения данного метода к созданию фрактальных фотонных структур.
	\item Узкий центральный максимум. Для проверки спектрального разрешения предлагаемого метода имеет смысл рассмотреть моделирование и экспериментальную реализацию узких максимумов, которые в общем случае достаточно трудно воспроизводимы, поскольку ширина максимума определяется минимальной шириной фотонной запрещённой зоны, достижимой для данного контраста показателей преломления.
\end{enumerate}
В связи с вышеизложенным, для демонстрации метода была выбрана функция именно такого вида.

\section{Алгоритм построения структуры фотонного кристалла}

Алгоритм строится следующим образом. Вначале выбирается желаемая функция спектрального отклика $f(\lambda)$. В рамках данной работы была выбрана функция, график которой представлен на рис. \ref{OriginalFunction}.

Выбиирается рабочий спектральный диапазон. Поскольку функция принимает ненулевые значения в диапазоне 475-725 нм, был выбран рабочий спектральный диапазон решения обратной задачи 450-750 нм. В этом диапазоне эквидистантно выбираются $N$ длин волн $\lambda_i$, которые соответствуют пространственным гармоникам.

Выбирается планируемая суммарная толщина фотоннокристаллической структуры $L$. Удобнее вести расчёт не в терминах физической толщины отдельного элемента $d$, а в терминах оптической толщины $nd$, поэтому здесь и далее в качестве пространственной координаты используется оптический путь. Суммарная толщина выбирается из соображений времени машинного счёта и технологических ограничений при изготовлении реальной структуры. Диапазон длины $L$ дискретизуется с шагом $\Delta x$, что соответствует физическому разделению структуры на слои $nd=\Delta x$. Выбор $L$ и $\Delta x$ фиксирует число слоёв структуры. В рамках данной работы выбрано значение $\Delta x = 20$ нм. Координата вдоль образца таким образом принимает дискретные значения $x_k=k\cdot \Delta x$, где $k$ - номер слоя.

Далее при помощи линейной интерполяции вычисляются значения функции $f(\lambda_i)$. Эти значения используются далее как амплитуды пространственных гармоник:

\begin{equation}
	r_i(x)=f(\lambda_i)\exp\left(i\dfrac{4\pi}{\lambda_i}x+\phi_i \right),
\end{equation}

где $x$ --- длина оптического пути в данной точке, $\phi_i$ --- фаза пространственной гармоники.

После этого выполняется суммирование всех пространственных гармоник и вычисление значения показателя преломления для каждого слоя фотонного кристалла:

\begin{equation}
	n_(x_k)=n_1+(n_2-n_1)\mathrm{Re}\left[\sum\limits_{i=1}^Nr_i(x_k)\right],
\end{equation}

где $n_1$ и $n_2$ --- минимальный и максимальный показатели преломления в рассматриваемой структуре. Таким образом, получается набор показателей преломления $n_k=n(x_k)$ для каждого слоя структуры.

Наконец, вычисляются значения физической толщины каждого слоя структуры:

\begin{equation}
d_k=\dfrac{\Delta x}{n_k}.
\end{equation}

Результатом работы алгоритма является набор значений $\{n_k,d_k\}$, который однозначно задаёт одномерную многослойную фотонную структуру.

\section{Используемые методы}
Для проверки работы алгоритма (обратная задача) выполнялось вычисление спектра коэффициента отражения (прямая задача). Задача вычисления спектра коэффициента отражения одномерной многослойной фотоннокристаллической структуры по известным значениям $\{n_k,d_k\}$ является известной и традиционно решается методом матриц распространения, описанным, например в \cite{luce2022tmm}.

Изготовление экспериментальных образцов фотонных кристаллов выполнялось при помощи электрохимического травления кремния, процедура детально описана в \cite{svyakhovskiy2012mesoporous}. Созданные по этому методу заготовки образцов из пористого кремния термически окислялись при температуре 700\textdegree С в течение 10 часов, в результате пористый кремний преобразовывался в пористый оксид кремния, или пористый плавленый кварц. В результате пористая, а следовательно, и фотоннокристаллическая структура образцов сохранялась, оптическое поглощение при этом уменьшалось. Детали процесса окисления и его влияние на оптические свойства ФК из пористого кремния также описаны в работе \cite{svyakhovskiy2012mesoporous}.

\section{Зависимость качества воспроизведения желаемой функции от толщины ФК}

\begin{figure}[tbh!]
	\centerline{\includegraphics[width=0.7\linewidth]{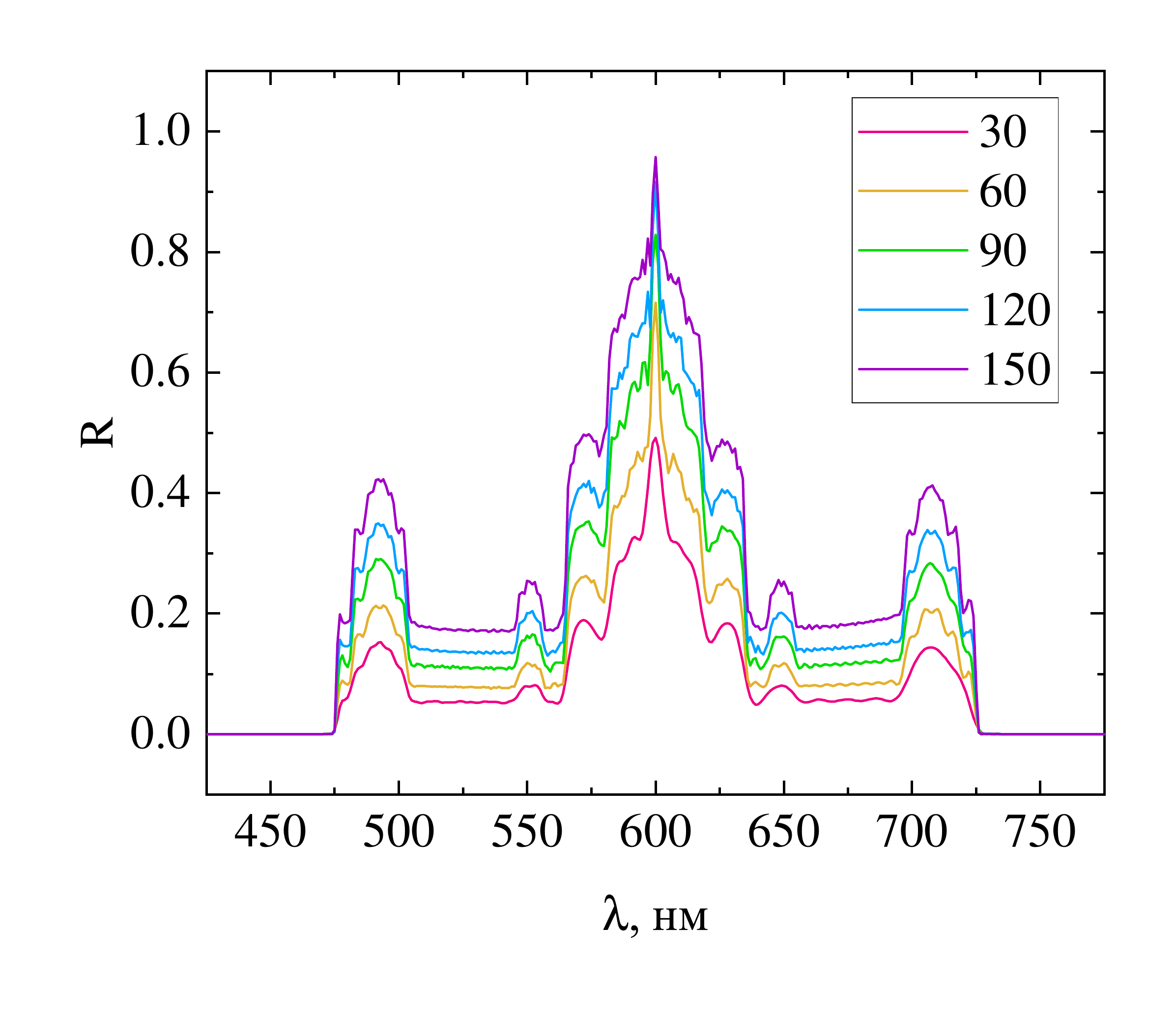}}
	\caption{Спектры коэффициента отражения фотонных структур, вычисленных по рассматриваемому алгоритму с ограничением суммарной оптической толщины структуры 30, 60, 90, 120, 150 мкм} \label{Theor_vs_l}
\end{figure}

Для выяснения минимально необходимой толщины фотоннокристаллической структуры был выполнен алгоритм построения структуры ФК с использованием различной оптической толшины структуры $L$, которая варьировалась от 30 до 150 микрон с шагом 30 микрон. На рис. \ref{Theor_vs_l} показаны спектры коэффициента отражения фотонных структур, полученных в результате работы алгоритма. В данном расчёте было использовано $N=1000$ пространственных гармоник. Диапазон показателей преломления был выбран от $n_1=1.14$ до $n_2=1.22$, что соответствует достижимым показателям преломления используемых в эксперименте фотонных кристаллов из пористого кварца.

Из результатов расчётов видно, что при малых толщинах ФК кривая спектрального отклика по форме похожа на желаемую функцию, однако имеет заниженную амплитуду, а тонкие спектральные особенности не прослеживаются. По мере увеличения толщины амплитуда отклика растёт и при значении 150 микрон становится похожей на желаемую функцию. Также постепенно проявляются спектральные особенности.

Этот результат объясняется тем, что собственные моды фотоннокристаллической структуры имеют конечную спектральную ширину, которая уменьшается по мере роста толщины структуры.

\begin{figure}[tbh!]
	\centerline{\includegraphics[width=0.7\linewidth]{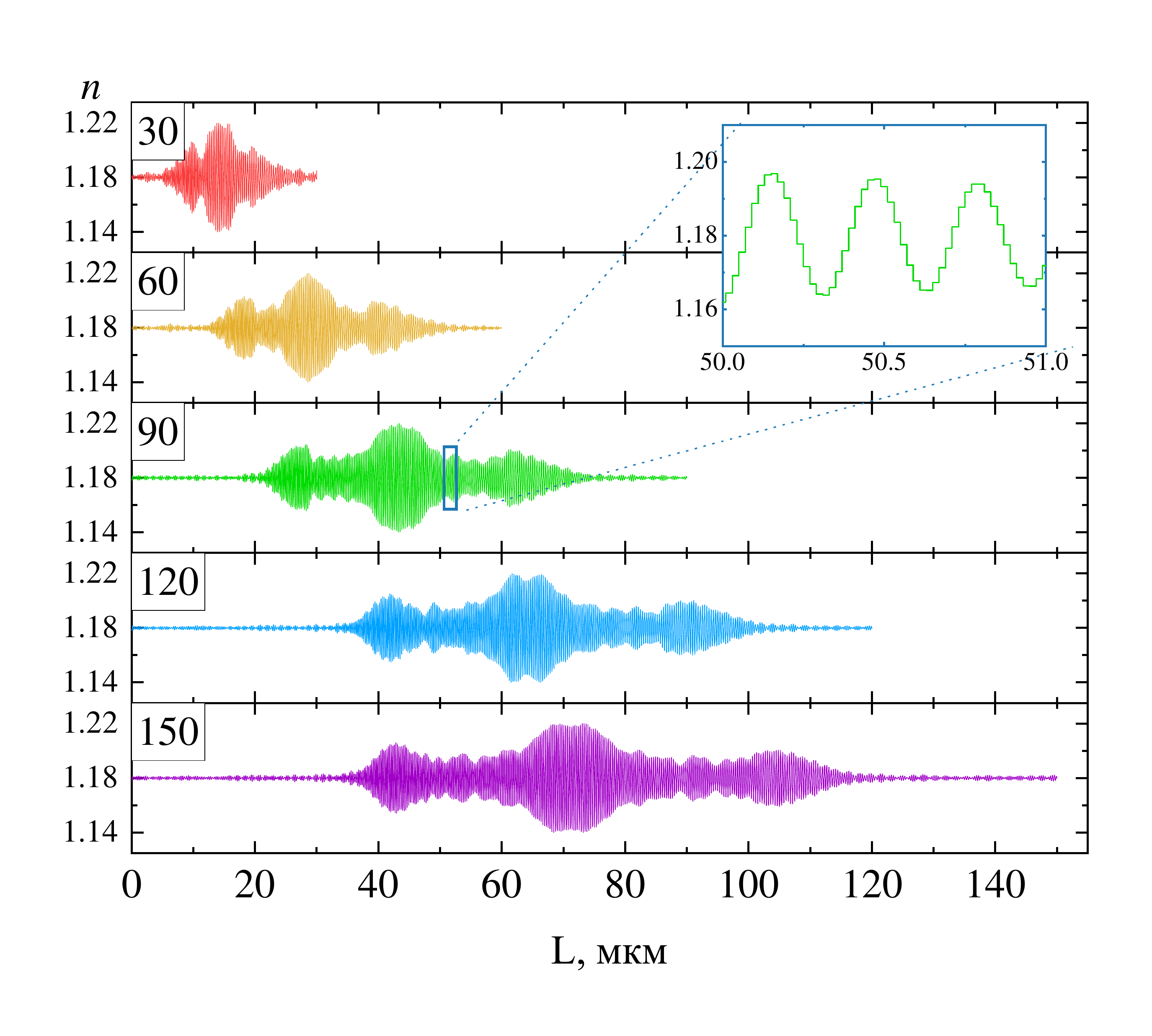}}
	\caption{Зависимость показателей преломления от длины оптического пути для фотонных структур, вычисленных по рассматриваемому алгоритму с ограничением суммарной оптической толщины структуры 30, 60, 90, 120, 150 мкм. Вставка: увеличенная часть структуры 90 мкм в диапазоне оптической толщины от 50 до 51 мкм} \label{Theor_nd}
\end{figure}

В целях более ясного понимания внутреннего строения фотоннокристаллической структуры на рис. \ref{Theor_nd} показаны явные зависимости показателя преломления от оптической толщины структуры. Показанные зависимости --- результат интерференции пространственных гармоник при их сложении.

Видно, что показатели преломления изменяются в диапазоне от $n_1=1.14$ до $n_2=1.22$, заметные части структуры на краях имеют показатель преломления, близкий к среднему значению 1.18. Средняя часть структуры имеет вид нескольких максимумов амплитуды модуляции, из которых можно выделить один главный и два побочных. При увеличении заданной оптической толщины средняя часть структуры растягивается в пространстве пропорционально суммарной толщине. На вставке показана увеличенная часть структуры, которая на малом пространственном масштабе имеет вид ступенчатой функции, аппроксимирующей модулированную синусоиду.

Отметим, что искусственное удаление краевых частей приводит к ухудшению спектральной функции и потере мелких деталей.

 В рамках этого раздела качество воспроизведения желаемой функции оценивалось по отсутствию паразитных осцилляций, было установлено, что паразитные осцилляции пропадают при увеличении толщины структуры до 150 мкм.
 
По результатам этого моделирования можно сделать вывод о том, что минимальная приемлемая оптическая толщина фотонной структуры для реализации желаемой функции составляет 150 мкм.

\section{Зависимость качества воспроизведения желаемой функции от числа пространственных гармоник}

\begin{figure}[tbh!]
	\centerline{\includegraphics[width=0.7\linewidth]{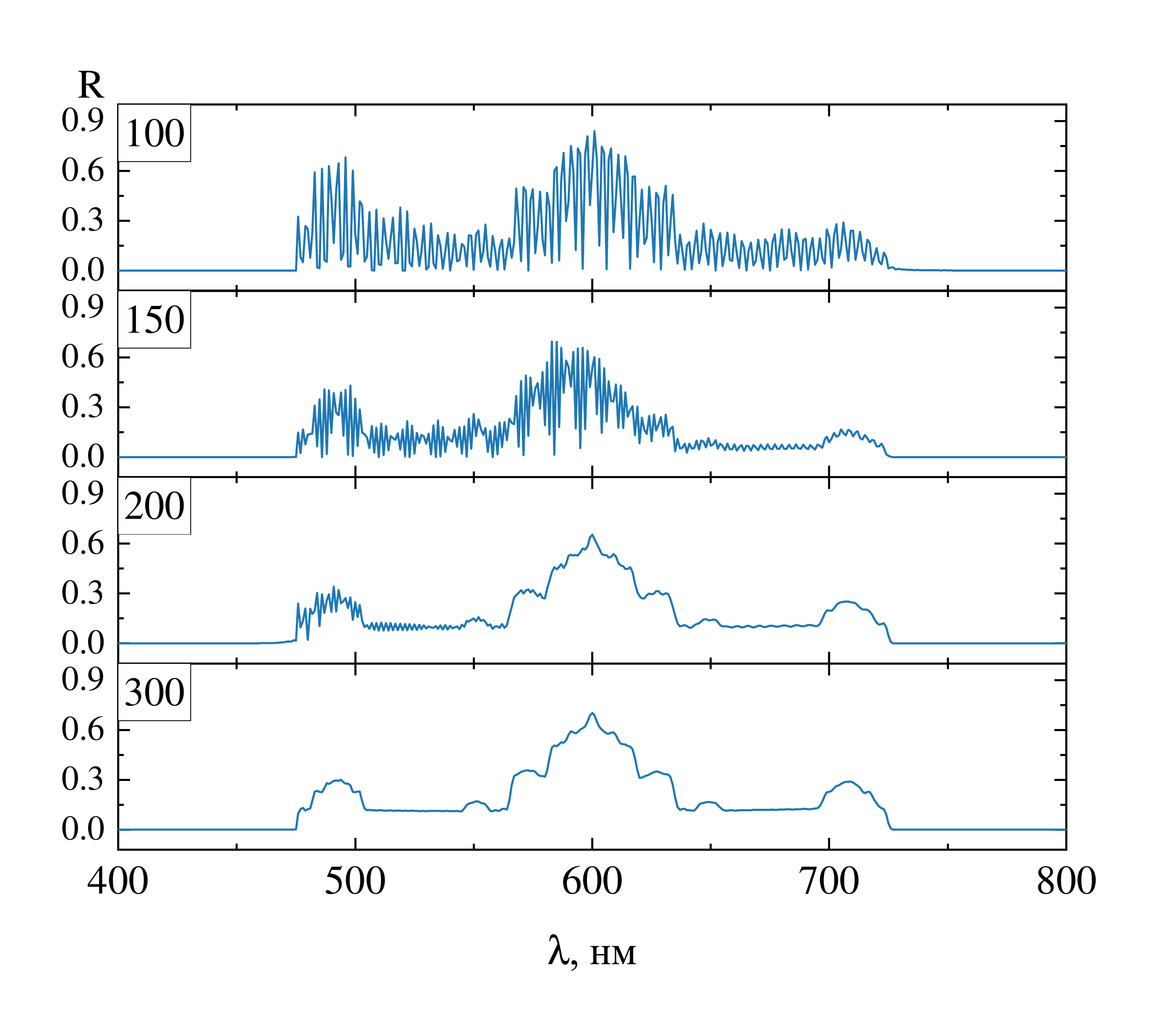}}
	\caption{Спектры коэффициента отражения фотонных структур, составленных для различного числа пространственных гармоник: 100, 150, 200, 300} \label{Theory_vs_N_100}
\end{figure}

В целях выяснения необходимого числа пространственных гармоник были построены ФК структуры при варьировании этого параметра. Для пространственной толщины 150 мкм были построены структуры, состоящие из 100, 150, 200, 300 гармоник. Соответствующие спектры коэффициента отражения этих структур показаны на рис. \ref{Theory_vs_N_100}.

Из моделирования видно, что при малом числе пространственных гармоник в спектре присутствуют шумы и паразитные осцилляции, которые исчезают по мере увеличения $N$. Вначале исчезают осцилляции в длинноволновой области, поскольку плотность собственных состояний ФК в этой области меньше, чем в коротковолновой, и поэтому требуется меньше пространственных гармоник для её качественной аппроксимации. 

\begin{figure}[tbh!]
	\centerline{\includegraphics[width=0.7\linewidth]{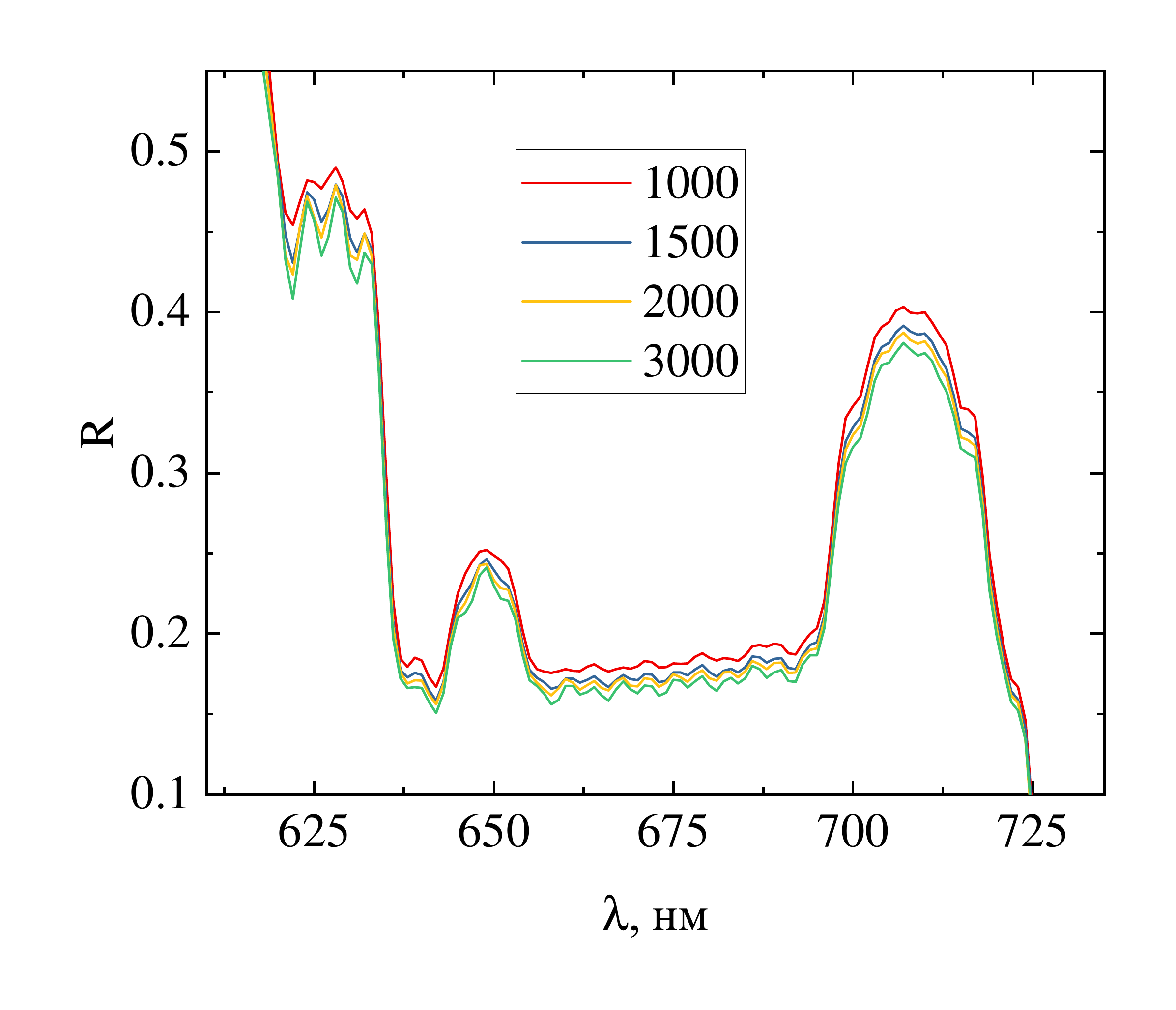}}
	\caption{Участок спектральной зависимости коэффициента отражения фотонных структур, составленных для различного числа пространственных гармоник: 1000, 1500, 2000, 3000} \label{Theory_vs_N_1000}
\end{figure}

Для воспроизведения мелких спектральных особенностей желаемой функции число пространственных гармоник было увеличено далее, на порядок. На рис.  \ref{Theory_vs_N_1000} показаны спектры коэффициента отражения структур, состоящих из 1000, 1500, 2000, 3000 гармоник соответственно. На рисунке представлен небольшой увеличенный фрагмент спектра от 610 до 740 нм, чтобы имелась возможность детально рассмотреть спектральные особенности.

Качество воспроизведения желаемой функции оценивалось по наличию в спектре малых спектральных особенностей, таких, как например группа пиков на длинах волн 629 и 632 нм, прямоугольные уступы на длинах волн 651, 701, 715 нм. Показано, что при увеличении числа гармоник до 3000 качество воспроизведения становится приемлемым, и это число гармоник было выбрано как оптимальное.

\section{Экспериментальные результаты и обсуждение}

\begin{figure}[tbh!]
	\centerline{\includegraphics[width=0.33\linewidth]{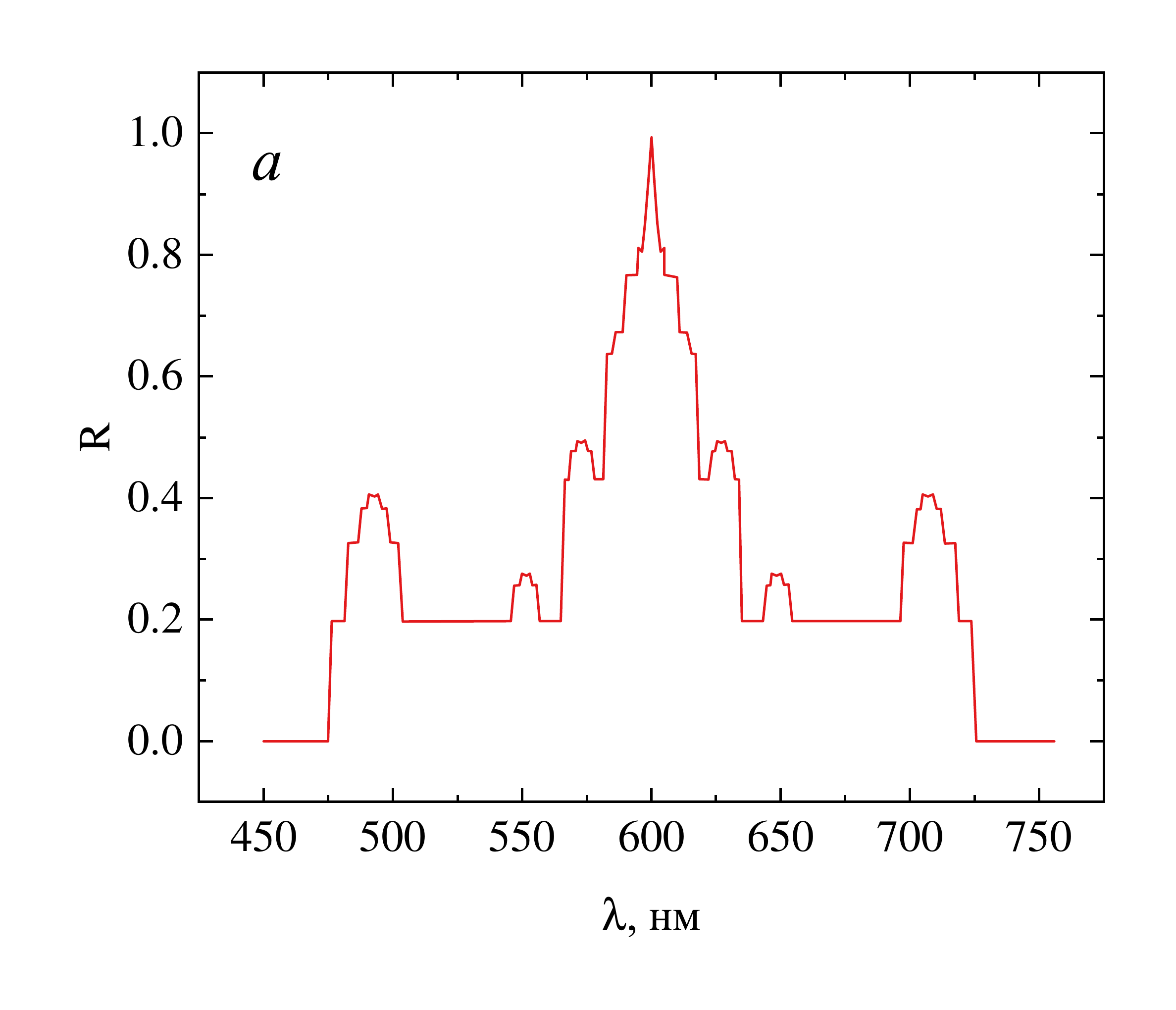}\includegraphics[width=0.33\linewidth]{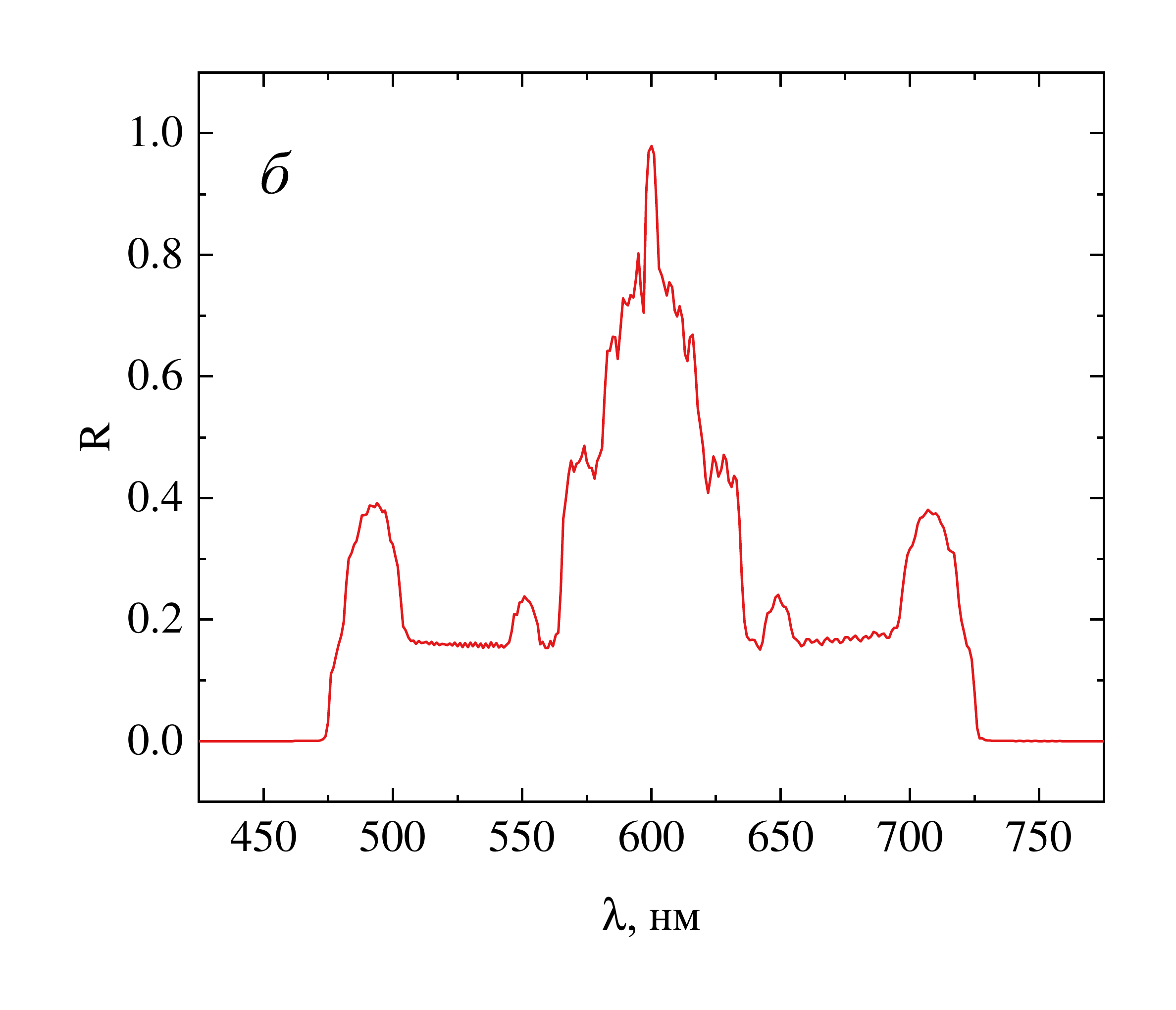}\includegraphics[width=0.33\linewidth]{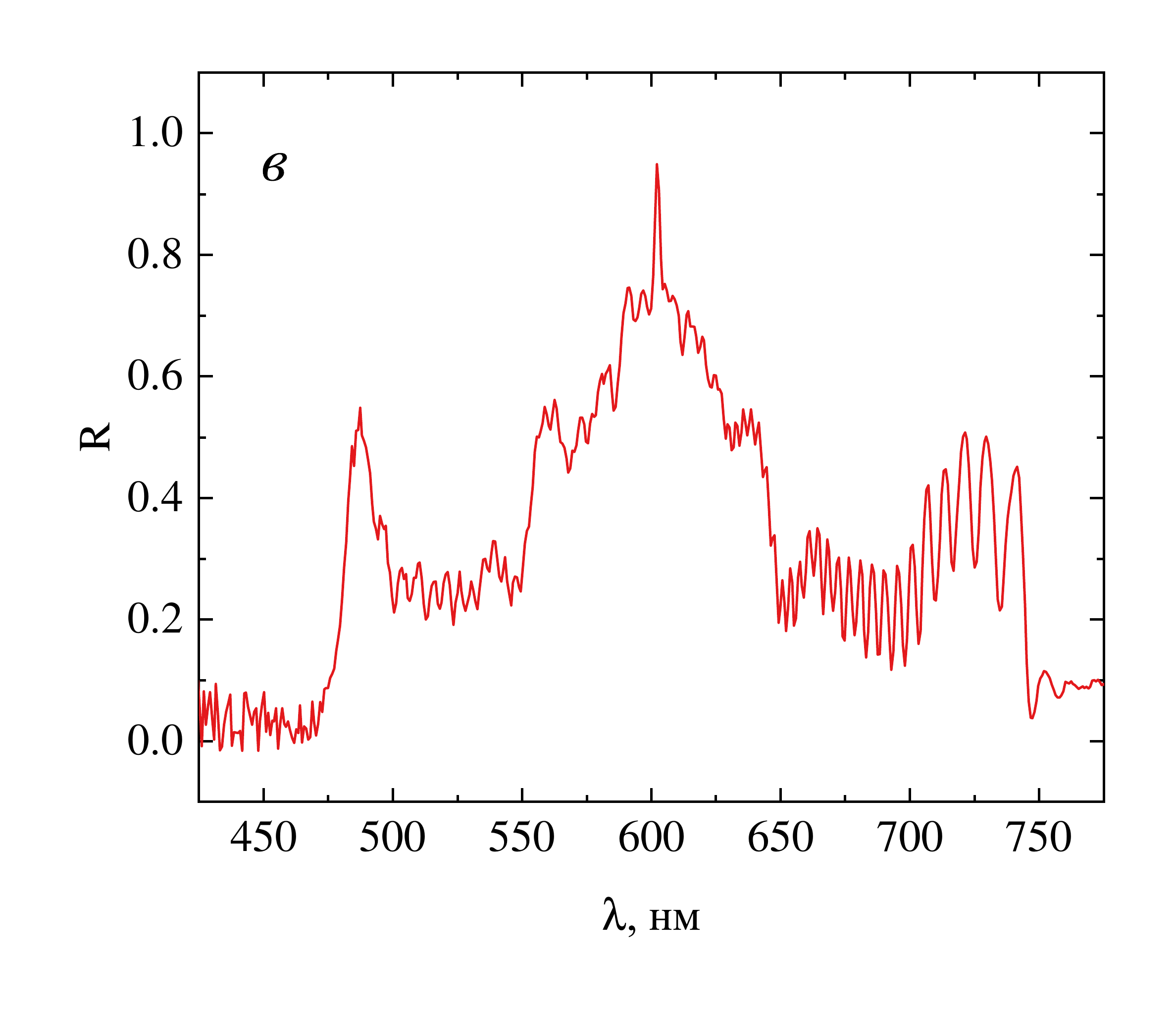}}
	\caption{Оптический спектр коэффициента отражения итоговой фотоннокристаллической структуры. \textit{а} --- желаемая функция, \textit{б} --- теоретический расчёт, \textit{в} --- эксперимент} \label{FinalExperiment}
\end{figure}

Итак, по итогам серий моделирования была выбрана минимально приемлемая толщина структуры $L=150$ мкм, число пространственных гармоник $N=3000$. Результат численного моделирования этой структуры в сравнении с желаемой функцией приведён на рис. \ref{FinalExperiment}б. В соответствии с рассчитанной структурой был экспериментально изготовлен фотонный кристалл, и измерен его спектр коэффициента отражения. Результат измерения показан на рис. \ref{FinalExperiment}в.

Видно, что теоретический расчёт хорошо аппроксимирует желаемую функцию. Экспериментально полученный спектр также находится в очень хорошем соответствии с функцией и расчётным спектром. А именно, воспроизведены П-образные участки с амплитудой, отличной от 1, и на этих участках кривая спектрального отклика имеет горизонтальную зависимость с уровнем $0.23\pm0.05$ (теоретическое значение 0.19). Наклонные участки треугольной формы повторяют форму желаемой функции, например, участок 604-612 нм имеет наклон -0.012 нм$^{-1}$ в расчёте и -0.0125 нм$^{-1}$ в эксперименте. Имеется хорошая детализация мелких спектральных особенностей, например, пиков 490 и 560 нм. Отчётливо виден центральный максимум, ширина которого составила $3.0\pm1$ нм (теоретическое значение 4 нм). Таким образом, воспроизведены все оговоренные выше спектральные особенности.

\section*{Заключение}
\mbox{}\vspace{-\baselineskip}
В работе продемонстрирован алгоритм построения фотоннокристаллической структуры, спектр которой имеет произвольно заданные особенности. Исследована применимость алгоритма при изменении толщины структуры и числа пространственных гармоник. Изложенный метод подтверждён экспериментально путём создания фотоннокристаллической структуры, спектр которого повторяет желаемую функцию. 

\bigskip
Авторы благодарят за финансовую поддержку этой работы Российский научный фонд, проект 21-72-10103, https://rscf.ru/project/21-72-10103/.

%\bibliographystyle{sse-vest}
%\bibliography{msu_spectrum}

\newpage
%КАЖДЫЙ РИСУНОК НА ОТДЕЛЬНОЙ СТРАНИЦЕ!!

\begin{flushright}
	\begin{tabular}{|ll|l|}
		\hline
		% after \\: \hline or \cline{col1-col2} \cline{col3-col4}...
		Рис 1 \mbox{}& &ВМУ. Физика. № \,\,-2023 \\
		к стр. \mbox{} & &К статье Свяховского С.Е.\\
		\hline
\end{tabular}\end{flushright}

\newpage
%КАЖДЫЙ РИСУНОК НА ОТДЕЛЬНОЙ СТРАНИЦЕ!!

\begin{flushright}
	\begin{tabular}{|ll|l|}
		\hline
		% after \\: \hline or \cline{col1-col2} \cline{col3-col4}...
		Рис 2 \mbox{}& &ВМУ. Физика. № \,\,-2023 \\
		к стр. \mbox{} & &К статье Свяховского С.Е.\\
		\hline
\end{tabular}\end{flushright}

\newpage
%КАЖДЫЙ РИСУНОК НА ОТДЕЛЬНОЙ СТРАНИЦЕ!!

\begin{flushright}
	\begin{tabular}{|ll|l|}
		\hline
		% after \\: \hline or \cline{col1-col2} \cline{col3-col4}...
		Рис 3 \mbox{}& &ВМУ. Физика. № \,\,-2023 \\
		к стр. \mbox{} & &К статье Свяховского С.Е.\\
		\hline
\end{tabular}\end{flushright}

\newpage
%КАЖДЫЙ РИСУНОК НА ОТДЕЛЬНОЙ СТРАНИЦЕ!!

\begin{flushright}
	\begin{tabular}{|ll|l|}
		\hline
		% after \\: \hline or \cline{col1-col2} \cline{col3-col4}...
		Рис 4 \mbox{}& &ВМУ. Физика. № \,\,-2023 \\
		к стр. \mbox{} & &К статье Свяховского С.Е.\\
		\hline
\end{tabular}\end{flushright}

\newpage
%КАЖДЫЙ РИСУНОК НА ОТДЕЛЬНОЙ СТРАНИЦЕ!!

\begin{flushright}
	\begin{tabular}{|ll|l|}
		\hline
		% after \\: \hline or \cline{col1-col2} \cline{col3-col4}...
		Рис 5 \mbox{}& &ВМУ. Физика. № \,\,-2023 \\
		к стр. \mbox{} & &К статье Свяховского С.Е.\\
		\hline
\end{tabular}\end{flushright}

\newpage
%КАЖДЫЙ РИСУНОК НА ОТДЕЛЬНОЙ СТРАНИЦЕ!!

\begin{flushright}
	\begin{tabular}{|ll|l|}
		\hline
		% after \\: \hline or \cline{col1-col2} \cline{col3-col4}...
		Рис 6 \mbox{}& &ВМУ. Физика. № \,\,-2023 \\
		к стр. \mbox{} & &К статье Свяховского С.Е.\\
		\hline
\end{tabular}\end{flushright}

\end{document}